\documentclass[fleqn,12pt,twoside]{article}
\usepackage{espcrc1}

\usepackage{graphicx}
\usepackage[T1]{fontenc}
\usepackage[latin1]{inputenc}
\usepackage{makeidx}
\usepackage{epsfig}
\usepackage{bm}
\usepackage{amsmath}
\usepackage{epstopdf}

\begin{document}

\title{Phase Transition in Small System}
\author{Ph. Chomaz$^{1},$ F. Gulminelli$^{2}$\thanks{%
member of the Institut Universitaire de France} \\
%EndAName
$^{1}$ GANIL, DSM-CEA/IN2P3-CNRS,\\
BP 5027, F-14076\ CAEN cedex 5, FRANCE\\
$^{2}$ LPC Caen, IN2P3-CNRS et Universit\'{e},\\
F-14050 CAEN cedex, FRANCE}
\maketitle

\begin{abstract}
Everybody knows that when a liquid is heated, its temperature increases
until the moment when it starts to boil. The increase in temperature then
stops, all heat being used to transform the liquid into vapor. What is the
microscopic origin of such a strange behavior? Does a liquid drop containing
only few molecules behave the same? Recent experimental and theoretical
developments seem to indicate that at the elementary level of very
small systems, this anomaly appears in an even more astonishing way: during
the change of state - for example from liquid to gas - the system cools
whereas it is heated, i.e. its temperature decreases while its energy
increases. This paper presents a review of our understanding of the
negative specific heat phenomenon.
\end{abstract}

\section{\protect\smallskip Introduction}

\smallskip Phase transitions are universal phenomena which have been
theoretically understood at the thermodynamic limit of infinite systems as
anomalies in the associated equation of state (EoS). They have been
classified according to the degree of non-analyticity of the thermodynamic
potential at the transition point. As an example figure 1 shows a first
order phase transition i.e. a jump in the average energy $<E>=-\partial 
_{\beta }$ $\log Z_{\beta },$ $Z_{\beta }$ being the canonical partition
sum, as a function of the conjugate intensive variable $\beta $.

\begin{figure}[tbp]
\begin{center}
\includegraphics[width=.5\textwidth]{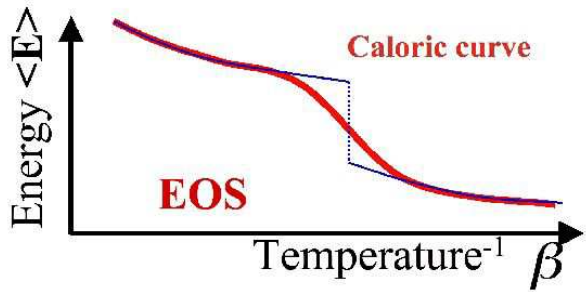}
\end{center}
\caption{Canonical caloric curve $<E>=-\partial _{\beta }\log Z_{\beta }$
presenting a discontinuity characteristic of a first order phase transition
for an infinite system (thin line) while it remains continuous in a finite
system (thick line). }
\label{fig:<E>beta}
\end{figure}

However, most of the systems studied in physics do not correspond
to this mathematical limit of infinite systems\cite{hill} and, in fact,
finite systems are now, per se, a subject of a very intense research
activity, from metallic clusters\cite{dauxois,clusters} to Bose condensates%
\cite{bose,traps}, from nanoscopic systems\cite{quantum} to atomic nuclei 
\cite{gross} and elementary particles\cite{qgp}.

\section{\protect\smallskip Finite systems}

The thermodynamic potentials of a finite system in a finite volume are
analytic functions. As a consequence, the average energy varies continuously
with the temperature and cannot present a jump. The transition looks
like a crossover (see fig. 1). Thus, it is often concluded that phase
transitions can only be defined for infinite systems.
This situation can be thought as unsatisfactory, since on one hand the
thermodynamic limit does not exist in nature, and on the other hand the
understanding of mesoscopic systems is becoming one of the most challenging
field in physics~\cite{wales}.

In the recent years, many efforts have been devoted to give the theoretical
foundations for the definition of phase transitions in small systems and
explicitly make the bridge between possible anomalies in finite systems and
the usual definition of phase transitions at the thermodynamical limit
%(infinite systems)
\cite{dauxois}.

\begin{figure}[tbp]
\begin{center}
\includegraphics[width=.5\textwidth]{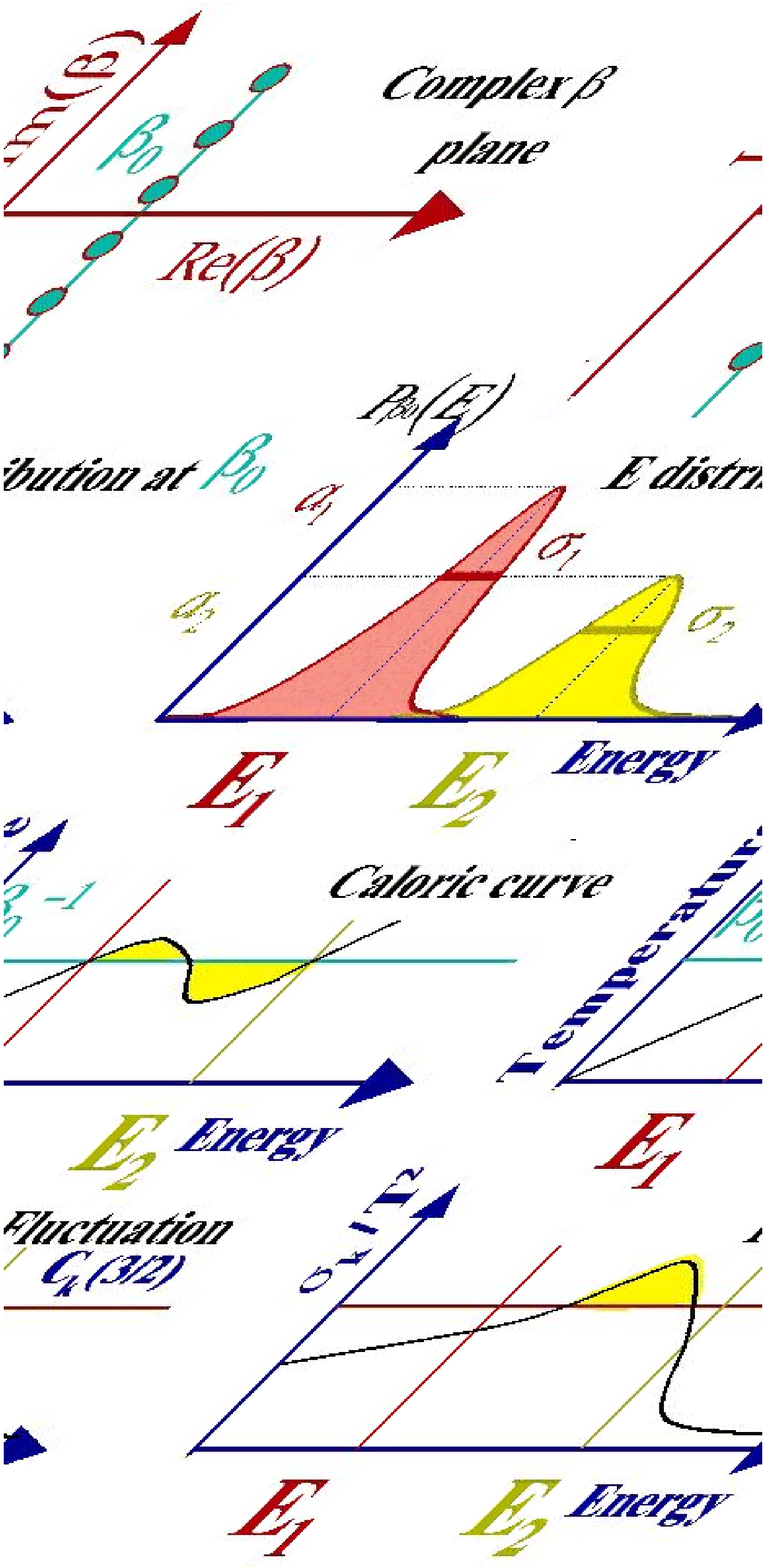}
\end{center}
\caption{Various equivalent definitions of a first order phase transition in
a finite system: an aligment of the zeroes of the partition sum in the
complex temperature plane (top); a bimodality of the event distribution
(below); a negative heat capacity (below) and an abnormal kinetic energy
fluctuation (bottom).}
\label{fig:theory}
\end{figure}

\subsection{\protect\smallskip Phase transition in the intensive ensembles:}

\textbf{Zeroes of }$\log Z$\textbf{\ and bimodality of the event
distribution.}

Following Yang and Lee\cite{yanglee} ideas, a classification scheme valid
for finite systems has been proposed by Grossmann~\cite{grossmann} using the
distribution of zeroes of the canonical partition sum in the complex
temperature plane (see top of fig. 2). Alternatively, we have recently
proposed the possible bimodality of the probability distribution of
observable quantities as a direct definition of a phase transition~\cite
{topology} (see fig. 2), the direction separating the two phases being an
order parameter.

A typical example, the distribution of the magnetization of the 3D Ising
model, is shown in figure 3. The observed %As a function of the temperature the
bifurcation is characteristic of a second order phase transition. Below the
critical point the bimodality in the magnetization transition indicates %the
%presence of 
a first order %phase 
transition at finite magnetic field (the
intensive variable conjugated to the magnetization).

\begin{figure}[tbp]
\begin{center}
\includegraphics[width=.5\textwidth]{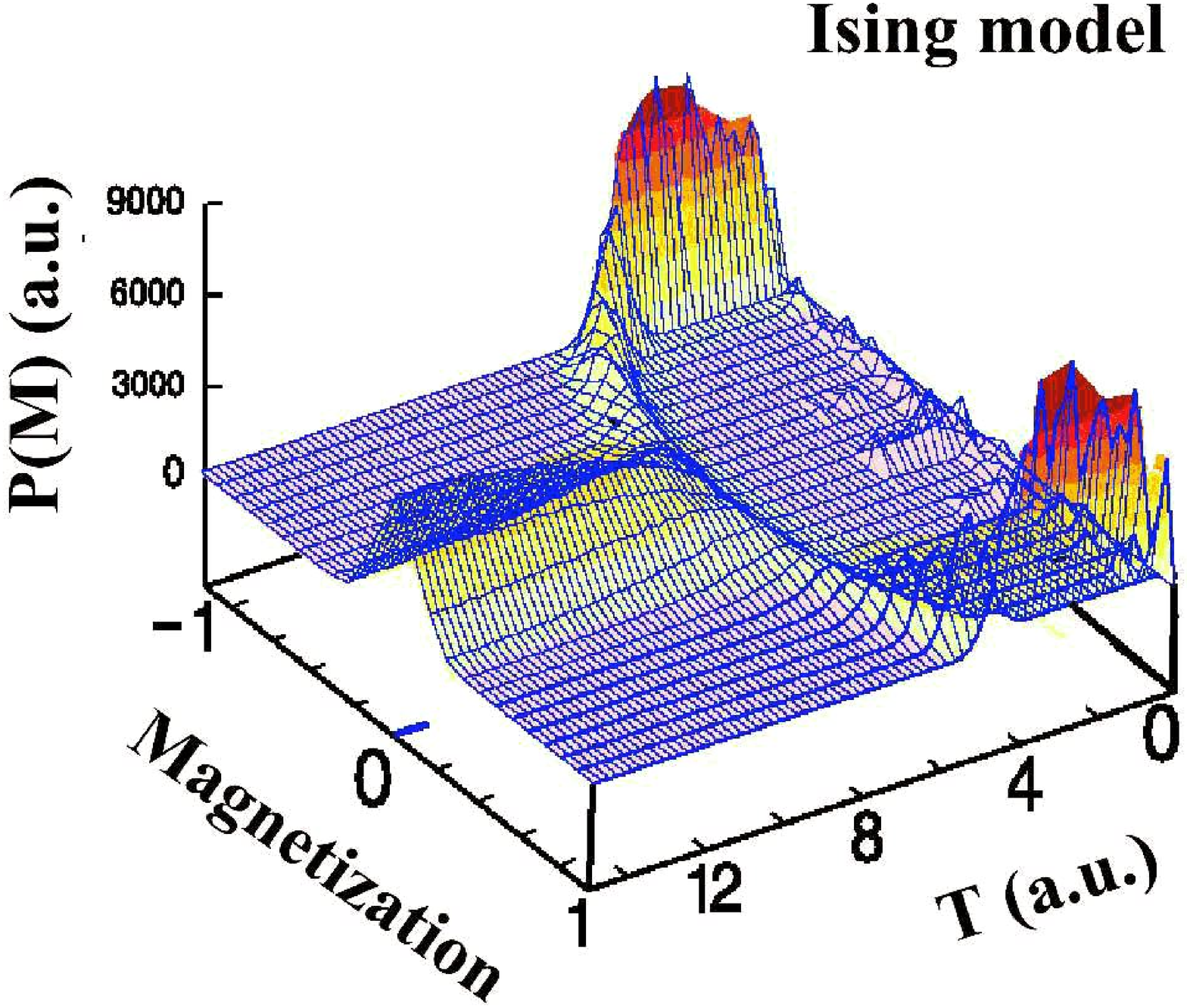}
\end{center}
\caption{Evolution of the magnetization distribution as a function of the
temperature in the Ising model ($\varepsilon /2$ is the strength of the spin
spin interaction) }
\label{Fig:Ising-M}
\end{figure}

In a recent work\cite{zeroes} we have demonstrate that this
definition of phase transitions in finite systems based on bimodality is
consistent with the Yang Lee theorem~\cite{yanglee}, $i.e.$ with the
standard definition of first order phase transitions in the thermodynamic
limit. The first step is to make the link between the partition sum and the
event distribution. Taking the example of a canonical ensemble both
quantities are directly related to the density of states $W(E):$ $P_{\beta
}(E)=W(E)e^{-\beta E}/Z_{\beta }$ and $Z_{\beta }=\int W(E)e^{-\beta E}dE.$
Therefore, introducing a reference temperature $\beta _{0}$ we get 
\begin{equation}
Z_{\beta }=Z_{\beta _{0}}\int P_{\beta _{0}}(E)e^{-(\beta -\beta _{0})E}dE
\label{EQ:Laplace-P}
\end{equation}
i.e. the partition sum is the Laplace transform of the event distribution.
%It should be noticed that 
A single distribution $P_{\beta
_{0}}(E)$ at a given temperature $\beta _{0}$) is in principle enough 
to reconstruct
the whole partition sum for all temperatures $\beta $. This is only true for
finite systems since at the thermodynamic limit the distribution $P$
becomes a delta function which contains information only about the %unique
temperature $\beta _{0}$ (as required by the theorem of ensembles equivalence).

Using (\ref{EQ:Laplace-P}) one can show that a set of zeroes converging and
aligning perpendicularly to the reel temperature axis, as required by the
Yang-Lee theorem, is equivalent to a probability distribution split at least
in 2 equivalent components (bimodal), separated by an energy inversely
proportional to the density of zeroes close to the real axis, $i.e.$
directly proportional to the number of particles $A$. 
%Such a density
%growing like $1/A$\ ($A$ the number of particles) is then equivalent to the
%persistence of a bimodal distribution with an energy difference proportional
%to $A.$ 
The resulting finite energy jump per particle is the latent heat.

\smallskip

\subsection{\protect\smallskip Link between intensive and extensive
ensembles:}

\textbf{Bimodality of the event distribution and negative heat capacity.}

Alternative definitions of phase transitions in finite systems have been
%recently 
proposed. It has been claimed that first order phase transitions in
finite systems can be related to a negative microcanonical heat capacity~%
\cite{gauss,gross} or more generally to an inverted curvature of the
thermodynamic potential as a function of an observable which can then be
seen as an order parameter~\cite{prl99}. These anomalies can also be
connected to the general topology of the potential energy surface~\cite
{pettini}.

It should be noticed that this definition of phase transitions can be used
only in\thinspace an ensemble in which an extensive variable\cite{warning} 
playing the role of an order parameter is controlled (extensive
ensemble) since it is based on the curvature anomalies of the associated
thermodynamical potential as a function of the controlled extensive
variable. Conversely, both the Yang-Lee and the bimodality pictures require
that the extensive variables %related to the order parameters 
are not
directly controlled but are %let 
free to fluctuate being constrained only by
a Lagrange multiplier
%\footnote{%
%which can be set to zero if needed in the case of no constraint.}
. This corresponds to the associated intensive ensemble. 
In this sense definitions
based on negative curvatures or on bimodalities (or on the zeroes of the
partition sum) are complementary, since they cannot be applied to the same
ensemble. 
%However, it is important to make the bridge between the two.

We have demonstrated that these two pictures are equivalent and are the
manifestation of the same phenomenon in two different ensembles\cite{zeroes}%
. This can be easily demonstrated. Let us take the energy as a typical
extensive variable. The inverted curvature definition of a first order phase
transition is nothing but the occurrence of a negative heat capacity in the
microcanonical ensemble, while the bimodality should be looked for in the
canonical energy distribution. However, using the definition of the
microcanonical entropy $S(E)=\log W(E)$ and of the canonical energy
distribution $P_{\beta }(E)$ one can directly show that they are intimately
related by the equality 
\begin{equation}
\log P_{\beta }(E)=S(E)-\beta E-\log Z_{\beta }  \label{EQ:P-S}
\end{equation}
%It should be noticed that 
The above relation is valid for every temperature $%
\beta ^{-1}$ and every energy $E$, meaning that in a finite system the energy
probability a single temperature is in principle enough to access the whole
entropy functional. However, from a practical point of view enough statistics
should be accumulated in a given energy bin in order to infer the associated
entropy so that Eq. (\ref{EQ:P-S}) can be used in practice only in the
strongly populated region.

From Eq. (\ref{EQ:P-S}) it is clear that the curvature of $\log P_{\beta }(E)
$ is directly the curvature of $S.$ Then, a bimodal probability distribution
implies a negative heat capacity and, vice versa, a negative entropy
curvature leads to a bimodal energy distribution for an canonical
temperature $\beta $ equal to the microcanonical one at the maximum of the
curvature inversion.

\begin{figure}[tbp]
\begin{center}
\includegraphics[width=.5\textwidth]{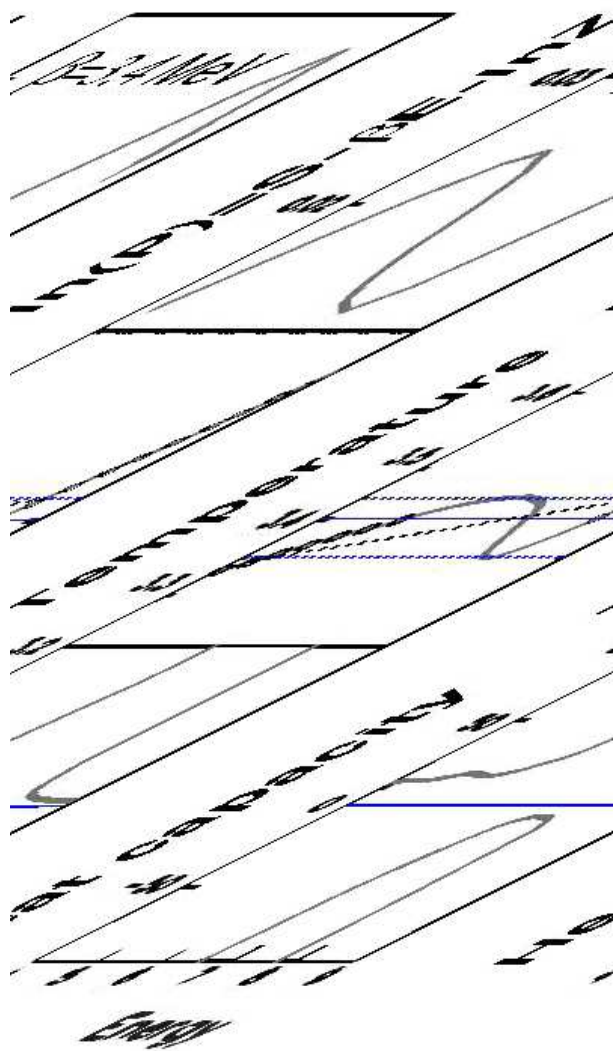}
\end{center}
\caption{Top: log of the energy distribution (a little above the transition
temperature) which is nothing but the entropy except for a subtracted linear
behavior. Middle: the deduced microcanonical temperature. For comparison the
canonical caloric curve $<E>$ as a funtion of $\beta ^{-1}$ as well as the
most probable energy are also plotted (dotted line and big dots
respectively). Bottom: the associated microcanonical heat capacity. }
\label{Fig:Lattice-gas-P-S-T-C}
\end{figure}

\smallskip An illustration of the relation between the signal of phase
transition in various ensembles in the case of an isobar Lattice-gas model%
\cite{isobar} is presented in figure \ref{Fig:Lattice-gas-P-S-T-C}. In the
top figure obtained in the canonical ensemble, the log of the energy
distributions is presented. The calculation
is performed close to the transition temperature, 
and the distribution is bimodal as expected. Because of eq. (\ref{EQ:P-S})
this figure also represents the entropy (with a linear behavior $%
\beta E+\log Z_{\beta }$ subtracted); its derivative is the
microcanonical temperature $T^{-1}=\partial _{E}S=\beta +$ $\partial 
_{E}P_{\beta }$ and its curvature the inverse of a heat capacity $C^{-1}=1/
\partial _{E}T=-T^{2}\partial _{EE}S=-T^{2}\partial _{EE}P_{\beta }.$ The
bimodality is nothing but a convex intruder in the entropy curve, which
produces a back bending in the microcanonical caloric curve and a negative
branch in the heat capacity (see figure \ref{Fig:Lattice-gas-P-S-T-C}).

For comparison on the microcanonical caloric curve of fig. (\ref
{Fig:Lattice-gas-P-S-T-C}) we also present the canonical average and most
probable energies. In the back-bending the canonical distribution is bimodal
so that the most probable energy is discontinuous (big dots) and so is
presenting the energy jump characteristic of a first order phase transition.
Conversely the usual canonical caloric curve which relates the average
energy to the temperature takes the whole back bending region %of temperature
(i.e. the whole range of temperatures exhibiting a bimodal energy
distribution) to go smoothly from the gas to the liquid (dotted line).

\subsection{\protect\smallskip Partitioning of energy and microcanonical
entropy}

\textbf{Negative heat capacity and abnormal kinetic energy fluctuations.}

In ref. \cite{npa99} we have proposed to use the fluctuations of the kinetic
energy to deduce the microcanonical heat capacity and we have shown that a
negative heat capacity results in an abnormally large energy fluctuation.
This can be easily explained for a classical fluid and tested in the
framework of the lattice-gas model. The total energy $E$ of the considered
system can be decomposed into two independent components, its kinetic and
potential energy: $E=K+V.$ In a microcanonical ensemble with a total energy $%
E$ the total degeneracy factor $W\left( E\right) =\exp \left( S\left(
E\right) \right) $ is thus simply given by the folding product of the
individual degeneracy factors $W_{k}\left( K\right) =\exp \left( S_{k}\left(
K\right) \right) $ and $W_{v}\left( V\right) =\exp \left( S_{v}\left(
V\right) \right) $. One can then define for the total system as well as for
the two subsystems the microcanonical temperatures $T_{i}$ and the
associated heat capacities $C_{i}$ (with $i=k$ or $v$). If we now look at
the kinetic energy distribution when the total energy is $E$ we get 
\index{microcanonical ensemble}

\begin{equation}
P_{E}\left( K\right) =\exp \left( S_{k}\left( K\right) +S_{v}\left(
E-K\right) -S\left( E\right) \right)  \label{eq:p1}
\end{equation}

Then the most probable kinetic energy $%
\bar{K}_{E}$ for the total energy $E$ is defined by the equality of the partial
microcanonical temperatures $T_{k}\left( \bar{K}_{E}\right) =T_{v}\left( E-
\bar{K}_{E}\right) $. The most probable kinetic energy $\bar{K}_{E}$ can be
used as a microcanonical thermometer. Using a Gaussian approximation for $
P_{E}\left( K\right) $ the kinetic energy variance can be calculated as \cite
{npa99}

\begin{equation}
\sigma _{k}^{2}=\bar{T}^{2}\frac{C_{k}C_{v}}{C_{k}+C_{v}}  \label{eq:sigma}
\end{equation}
where $C_{k}$ and $C_{v}$ are the microcanonical heat capacities calculated
for the most probable energy partition. Equation (\ref{eq:sigma}) can be
inverted to extract from the observed fluctuations the heat capacity 
\index{anomalous fluctuations}

\begin{equation}
C\simeq C_{k}+C_{v}=%
\frac{C_{k}}{1-\sigma _{k}^{2}/\bar{\sigma}_{k}^{2}}  \label{EQ:14}
\end{equation}
where we have introduced the canonical kinetic energy fluctuation $\bar{
\sigma}_{k}^{2}=\bar{T}^{2}C_{k}.$ This expression shows that when $C_{v}$
diverges and then becomes negative, $\sigma _{k}^{2}$ remains positive but
overcomes the canonical expectation $\bar{\sigma}_{k}^{2}=\bar{T}^{2}C_{k}$.
This anomalously large kinetic energy fluctuation is a signature of the
first order phase transition.

\begin{figure}[tbp]
\begin{center}
\includegraphics[width=.9\textwidth]{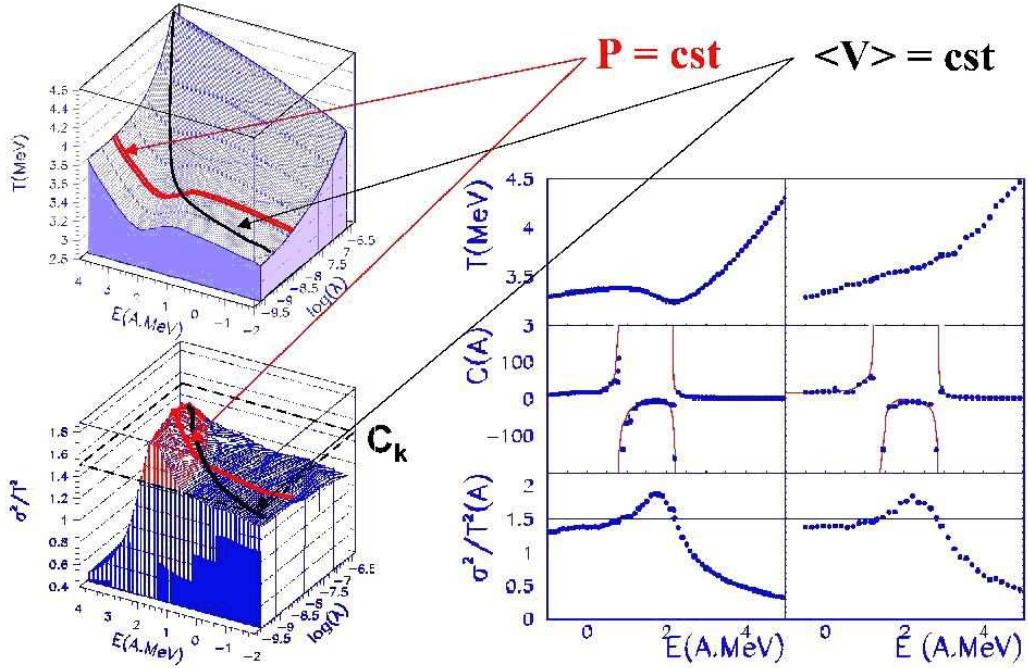}
\end{center}
\caption{Isobar lattice gas model: Left part : Isotherms and contour plot of
the normalized kinetic energy fluctuations in the Lagrange parameter versus
energy plane. The level corresponding to the canonical expectation $\sigma
_{1}^{2}/T^{2}=1.5$ is shown. Thick line: critical isotherm. Right panel :
Thermodynamic quantities in the microcanonical ensemble for a
transformation at constant pressure and at constant volume (right part).
Upper panels: caloric curve. Lower panels: normalized kinetic energy
fluctuations compared to the canonical expectation (lines). Medium panels:
heat capacity (symbols) compared to the estimation through eq.(5) (lines) at
constant pressure (left part) and at constant volume (right part). }
\label{fig_bologne4}
\end{figure}

The normalized fluctuations $\sigma _{K}^{2}/${$T_{\lambda }$}$^{2}$
obtained in the microcanonical ensemble for a lattice-gas with an average
volume constrained by a Lagrange multiplier $\lambda $ are shown in the
energy-$\lambda $ plane in figure  \ref{fig_bologne4} together with the
isotherms. One can clearly see that the
fluctuations are abnormally large in the coexistence region
up to the critical temperature. From fig. 
\ref{fig_bologne4} it is apparent that the phase transition signal is
visible in the temperature (back-bending) as well as in the fluctuation
observable (normalized fluctuations larger than $C_{k})$.

The right part of the figure presents examples of expected behavior of the
temperature as a function of energy at a constant pressure or a constant
average volume in the subcritical region. At constant pressure 
%the caloric curves are
%steeper than the ones at constant $\lambda $ when
%the system is in the liquid or in the vapor phase; in the coexistence region
%the isobars are almost identical to the iso-$\lambda $'s since $P_{\lambda }$
%and $\lambda $ differ only by the temperature which is almost constant in
%the phase transition region, and 
a back-bending is clearly seen. On the other hand at constant average volume
a smooth behavior is observed with a slope change entering the gas phase.
This is due to the fact that the $\lambda $ parameter varies rapidly in the
coexistence region. From these examples one clearly sees that the various
transformations lead to very different caloric curves. More generally, it is
clear that the back-bending of the temperature surface can be avoided
depending on the path of the considered transformation and the phase
transition signal can be hidden in the observation of the caloric curve.

On the other side partial energy fluctuations are a state variable which
does not depend on the transformation from one state to another and can
directly give access to the equation of state: 
%From figure \ref{fig_bologne4}
%we can see that in the whole phase transition region the microcanonical
%fluctuations present a strong maximum which exceeds the canonical value: 
the anomalously large fluctuation signal will be always seen if the system
undergoes a first order phase transition, independent of the path. As an
example the lower part of figure 5 shows a constant $P_{\lambda }$ or $%
<V>_{\lambda }$ cut of the bidimensional fluctuation surface. The
quantitative behavior of the heat capacity as a function of energy depends
on the specific transformation, but at each point the heat capacity
extracted from fluctuations is a direct measure of the underlying equation
of state. This is clearly demonstrated in the medium part of figure \ref
{fig_bologne4} which compares the exact heat capacity $C_{\lambda }$ with
the fluctuation approximation. The agreement between the two results
illustrates the accuracy of the estimation (14).

\section{ Conclusions}

Phase transitions are universal properties of matter in interaction. They
have been widely studied in the thermodynamical limit of infinite systems.
However, in many physical situations this limit cannot be accessed and so
phase transitions should be reconsidered from a more general point of view.
This is for example the case of matter under long range forces like
gravitation. Even if these self gravitating systems are very large they
cannot be considered as infinite because of the non saturating nature of the
force. Other cases are provided by microscopic or mesoscopic systems 
built out of matter which is known to present phase transitions. Metallic
clusters can melt before being vaporized. Quantum fluid may undergo Bose
condensation or super-fluid phase transition. Dense hadronic matter should
merge in a quark and gluon plasma phase while nuclei are expected to exhibit
a liquid -gas phase transition. For all these systems the theoretical and
experimental issue is how to sign a possible phase transition in a finite
system.

In this paper we have shown that phase transitions can be uniquely defined
for finite systems. Depending upon the statistical ensemble one should look
for different signals. In the ensemble where the order parameter is free to
fluctuate (intensive ensemble) %we have shown that 
the topology of the event
distribution should be studied. A bimodal distribution signals a first order
phase transition. This occurrence of a bimodal distribution is equivalent
to the alignment of the partition sum zeroes as described by the Yang and Lee
theorem.
In the associated extensive ensemble, the bimodality condition is equivalent
to the requirement of a convexity anomaly in the thermodynamic potential. 

The first experimental evidences of such a phenomenon have been reported
recently different fields:

the melting of sodium clusters\cite{haberland}, the fragmentation of
hydrogen clusters\cite{Farizon}, the pairing in nuclei\cite{Melby} and 
nuclear multifragmentation\cite{Prague}. 
However, much more experimental and theoretical
studies are now expected to progress in this new field of phase transitions
in finite systems.


\begin{thebibliography}{99}
\bibitem{warning}
In this paper the word ''extensive'' is used in the general sens of
resulting from an observation and not in the restricted sense of additive
variable. Intensive variables are conjugate to extensive variables i.e.
Lagrange multipliers imposing the average value of the associated extensive
variable.
\end{thebibliography}
\end{document}